\documentclass[12pt]{article}

\usepackage[margin=1in]{geometry}
\usepackage{amsmath} 

\usepackage{graphicx}    % for figures
\usepackage{listings}    % for code blocks
\usepackage{booktabs}    % for table rules
\usepackage{array}       % for table column alignment if needed
\usepackage{xspace}      % optional, helps spacing around macros if needed
\usepackage{url}         % for URLs in the bibliography
\usepackage{float}       % for floating tables/figures if needed

\title{AbsInf: A Lightweight Object to Represent \texttt{float('inf')} in Dijkstra’s Algorithm}

\author{
  Anjan Bellamkonda \\
  Independent Researcher \\
  Undergraduate Student, \\
  College of Engineering, Virginia Tech \\
  (Unaffiliated) \\
  \texttt{anjanbellamkonda@vt.edu}
  \and
  Laksh Bharani \\
  Independent Researcher \\
  Undergraduate Student, \\
  College of Engineering, Virginia Tech \\
  (Unaffiliated) \\
  \texttt{lakshb2005@vt.edu}
  \and
  Harivatsan Selvam \\
  Independent Researcher \\
  Alpharetta High School, GA \\
  (Unaffiliated) \\
  \texttt{sharivats2007@gmail.com}
}

\date{}

\begin{document}
\maketitle

\begin{abstract}
We introduce \textbf{AbsInf}, a lightweight abstract object designed as a high-performance alternative to Python’s native \texttt{float('inf')} within pathfinding algorithms. Implemented as a C-based Python extension, AbsInf bypasses IEEE-754 float coercion and dynamic type dispatch, offering constant-time dominance comparisons and arithmetic neutrality. When integrated into Dijkstra’s algorithm without altering its logic, AbsInf reduces runtime by up to 17.2\%, averaging 9.74\% across diverse synthetic and real-world graph datasets. This optimization highlights the performance trade-offs in high-frequency algorithmic constructs, where a symbolic use of infinity permits efficient abstraction. Our findings contribute to the broader discourse on lightweight architectural enhancements for interpreted languages, particularly in performance-critical control flows.
\end{abstract}

\section{Introduction}

Graph algorithms are foundational in modern computing, with applications spanning network routing, dependency resolution, and a variety of optimization problems. Dijkstra’s algorithm, in particular, remains a widely studied shortest-path solution, especially in Python because of its readability and accessibility~\cite{ref2}. A common pattern in such algorithms is the initialization of node distances using Python’s built-in \texttt{float('inf')}, which represents positive infinity in the IEEE-754 floating-point standard.

While Python’s native \texttt{float('inf')} is functionally sufficient, it introduces performance overhead in tight loops. Each use incurs standard floating-point comparison and arithmetic costs, and Python’s dynamic type system promotes values to \texttt{float} before evaluating expressions like \(\texttt{x} < \texttt{float('inf')}\). These operations, though fast in isolation, can become non-trivial in large-scale, high-frequency scenarios.

A new version of IEEE 754 was published in 2008, following a multi-year revision process~\cite{ref4}. The current IEEE~754--2019 standard includes clarifications and new operations but retains the core binary/decimal formats. Python’s native \texttt{float('inf')} implements a binary64 format~\cite{ref5}, which can be overkill for algorithms that only need a symbolic representation of infinity. For instance, the binary64 format stores a 64-bit representation, meaning the CPU must process all 64 bits even for a simple comparison.

\begin{figure}[H]
\centering
\includegraphics[width=0.6\linewidth]{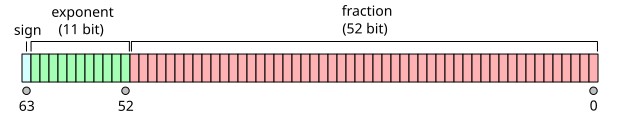}
\caption{Representation of infinity in a binary64 float, consisting of sign, exponent, and fraction bits~\cite{ref3}.}
\end{figure}

To address this, we introduce \textbf{AbsInf}, a custom Python object implemented as a lightweight C extension. AbsInf mimics the behavior of \texttt{float('inf')} in algorithmic contexts but is designed explicitly for speed, where comparisons like \texttt{AbsInf} \(>\) \texttt{x} always return \texttt{True} in constant time, and arithmetic operations like \(\texttt{AbsInf} + \texttt{x}\) skip floating-point computation entirely. By directly handling the low-level behavior, AbsInf eliminates numeric coercion and float-level branching associated with Python’s default infinity representation.

We integrate AbsInf into Dijkstra’s algorithm by replacing only the distance-initialization step, leaving the rest of the logic unchanged. Benchmarks spanning ten distinct synthetic graph structures and eight real-world routes demonstrate consistent runtime gains, with peak improvements of 17.2\% and an overall average of 9.74\%. Our findings highlight that even small, domain-specific optimizations can yield notable performance benefits in high-level languages like Python.

\section{Methodologies}

\subsection{Object Behavior}

AbsInf is designed to mimic three core behaviors of \texttt{float('inf')}:

\begin{enumerate}
\item \textbf{Comparison Dominance:} \texttt{AbsInf} should always be treated as greater than any numeric value.
\item \textbf{Addition Neutrality:} \(\texttt{AbsInf} + \texttt{x}\) should return \texttt{AbsInf}.
\item \textbf{Minimal Overhead:} All operations should complete without invoking Python’s float coercion or floating-point arithmetic pathways.
\end{enumerate}

These behaviors are implemented at the C level to bypass Python’s dynamic dispatch. The code snippet below illustrates the core logic:

\begin{lstlisting}[language=C, basicstyle=\small\ttfamily]
#define PY_SSIZE_T_CLEAN
#include <Python.h>

typedef struct {
    PyObject_HEAD
} AbsInfObject;

static PyObject *
AbsInf_new(PyTypeObject *type, PyObject *args, PyObject *kwds) {
    return type->tp_alloc(type, 0);
}

static PyObject *
AbsInf_richcompare(PyObject *self, PyObject *other, int op) {
    switch (op) {
        case Py_GT:
            Py_RETURN_TRUE;
        case Py_LT:
            Py_RETURN_FALSE;
        default:
            Py_RETURN_NOTIMPLEMENTED;
    }
}

static PyObject *
AbsInf_add(PyObject *self, PyObject *other) {
    Py_INCREF(self);
    return self;
}

static PyNumberMethods AbsInf_as_number = {
    .nb_add = AbsInf_add,
};

static PyTypeObject AbsInfType = {
    PyVarObject_HEAD_INIT(NULL, 0)
    .tp_name = "absinf.AbsInf",
    .tp_basicsize = sizeof(AbsInfObject),
    .tp_itemsize = 0,
    .tp_flags = Py_TPFLAGS_DEFAULT,
    .tp_new = AbsInf_new,
    .tp_richcompare = AbsInf_richcompare,
    .tp_as_number = &AbsInf_as_number,
    .tp_doc = "AbsInf: always greater than any number",
};

static PyModuleDef absinfmodule = {
    PyModuleDef_HEAD_INIT,
    .m_name = "absinf",
    .m_doc = "Abstract infinity object that is greater than any number.",
    .m_size = -1,
};

PyMODINIT_FUNC
PyInit_absinf(void) {
    PyObject *m;
    if (PyType_Ready(&AbsInfType) < 0)
        return NULL;

    m = PyModule_Create(&absinfmodule);
    if (m == NULL)
        return NULL;

    Py_INCREF(&AbsInfType);
    if (PyModule_AddObject(m, "AbsInf", (PyObject *)&AbsInfType) < 0) {
        Py_DECREF(&AbsInfType);
        Py_DECREF(m);
        return NULL;
    }

    return m;
}
\end{lstlisting}

\subsection{Integration with Dijkstra’s Algorithm}

To integrate AbsInf, we replace the default initialization of distances (i.e., \texttt{float('inf')}) with \texttt{AbsInf()}. The remainder of Dijkstra’s logic (e.g., picking the smallest distance, updating neighbors) remains unchanged:

\begin{lstlisting}[language=Python, basicstyle=\small\ttfamily]
def dijkstra_basic(graph, start):
    unvisited = list(graph.keys())
    distances = {node: float('inf') for node in graph}
    distances[start] = 0

    while unvisited:
        current = min((node for node in unvisited), key=lambda node: distances[node])
        for neighbor, weight in graph[current].items():
            new_distance = distances[current] + weight
            if distances[neighbor] < new_distance:
                distances[neighbor] = new_distance
        unvisited.remove(current)

    return distances

def dijkstra_abs(graph, start):
    from absinf import AbsInf
    unvisited = list(graph.keys())
    distances = {node: AbsInf() for node in graph}
    distances[start] = 0

    while unvisited:
        current = min((node for node in unvisited), key=lambda node: distances[node])
        for neighbor, weight in graph[current].items():
            new_distance = distances[current] + weight
            if distances[neighbor] < new_distance:
                distances[neighbor] = new_distance
        unvisited.remove(current)

    return distances
\end{lstlisting}

\subsection{Testing Environments}

We tested both implementations (\texttt{float('inf')} and AbsInf) on ten synthetic graph categories, each repeated twice for statistical smoothing. Graph types ranged from sparse trees and dense graphs to cycle and disconnected structures. The performance was measured on a machine with an Intel\textsuperscript{\textregistered}~Core\texttrademark~Ultra 9 185H CPU (2.30\,GHz, 16~cores), 32\,GB RAM, running Windows~11 Education 24H2. All tests were conducted using CPython 3.12.0 (64-bit), and timings were collected with Python’s built-in \texttt{timeit} module over 50{,}000 iterations. The \texttt{AbsInf} object was compiled as a native C extension via \texttt{setuptools}.

For real-world routing tests, we used eight different route graphs (downloaded from OpenStreetMap), each repeated 10 times on a Microsoft Azure server (Standard E2s v3, 2 vCPUs, 16\,GiB memory) running Windows Server~2025 Datacenter OS with Python~3.13.

\subsection{Test Cases}

The ten synthetic graph examples and the eight real-world routes are detailed in the Appendix, along with the Python scripts used to benchmark each scenario.

\subsection{Welch’s T-Test}

\noindent
\textbf{Objective:} Determine whether the mean runtime of Dijkstra’s algorithm with \texttt{AbsInf} is significantly less than that with \texttt{float('inf')}.

\begin{itemize}
\item \textbf{Null Hypothesis} (\(H_0\)): \(\mu_{\text{AbsInf}} \ge \mu_{\text{float-inf}}\).
\item \textbf{Alternative Hypothesis} (\(H_a\)): \(\mu_{\text{AbsInf}} < \mu_{\text{float-inf}}\).
\end{itemize}

We used a one-tailed Welch’s t-test (assuming unequal variances). With a sample size \(n=40\), normal approximation is reasonable. The significance level was set to \(\alpha=0.01\).

\section{Results}

\subsection{Synthetic Maps: Average Runtime Improvements}

\begin{table}[H]
\centering
\begin{tabular}{lccc}
\toprule
\textbf{Graph Type} & \textbf{float('inf')} & \textbf{AbsInf} & \(\%\) \textbf{Improvement} \\
\midrule
Linear Chain        & 0.1874   & 0.1647   & \(\sim\)12.1\% \\
Sparse Tree         & 0.2378   & 0.1968   & \(\sim\)17.2\% \\
Dense Graph         & 0.1863   & 0.1759   & \(\sim\)5.6\%  \\
Star Graph          & 0.1847   & 0.1788   & \(\sim\)3.2\%  \\
Disconnected Graph  & 0.2051   & 0.1883   & \(\sim\)8.2\%  \\
Cycle Graph         & 0.1431   & 0.1269   & \(\sim\)11.3\% \\
Equal Weights       & 0.1597   & 0.1513   & \(\sim\)5.3\%  \\
Large Uniform Graph & 0.4768   & 0.4391   & \(\sim\)7.9\%  \\
Worst-Case Tie      & 0.1677   & 0.1440   & \(\sim\)14.2\% \\
Real-World-Like     & 0.1976   & 0.1727   & \(\sim\)12.6\% \\
\bottomrule
\end{tabular}
\caption{Runtime improvements for different categories of synthetic maps (average over 50,000 iterations).}
\end{table}

\begin{figure}[H]
\centering
\includegraphics[width=0.6\linewidth]{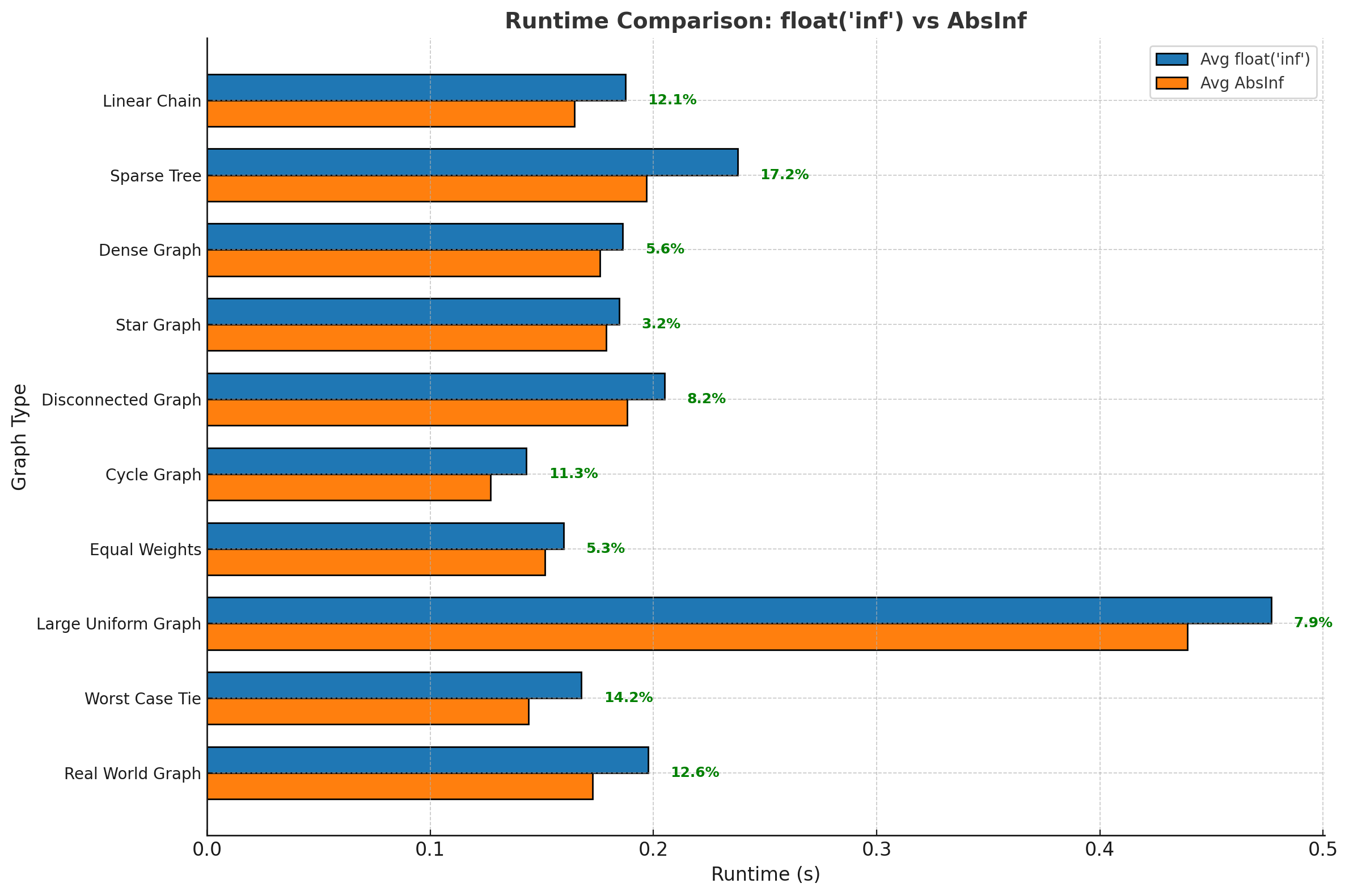}
\caption{Bar chart indicating the runtimes and percent improvements across synthetic map types.}
\end{figure}

\subsection{Statistical Significance (Welch’s T-Test)}

The one-tailed p-value for the difference in mean runtimes was \(0.0002175\). Since \(p<0.01\), we reject \(H_0\) at the 99\% confidence level, concluding that Dijkstra’s algorithm with AbsInf is significantly faster than with \texttt{float('inf')}.

\subsection{Real-World Routes: Sample Improvements}

\begin{table}[H]
\centering
\begin{tabular}{lccc}
\toprule
\textbf{Route} & \textbf{float('inf')} & \textbf{AbsInf} & \(\%\) \textbf{Improvement} \\
\midrule
McComas Hall \(\to\) Kroger South      & 2.0755   & 1.8738   & 9.72           \\
McComas Hall \(\to\) Kroger North      & 2.0768   & 1.8584   & 10.52          \\
The Inn \(\to\) Pamplin Hall           & 0.7183   & 0.6942   & 3.36           \\
Fairfield Marriott \(\to\) Decathlon   & 3277.814 & 2764.523 & 15.66          \\
St. Johns Bus Stand \(\to\) Rameshwaram Caf\'{e} & 10125.57 & 8979.323 & 11.32 \\
M.A. Chidambaram Stadium \(\to\) Chennai Lighthouse & 46.4325 & 41.6563 & 10.29 \\
Dr. MGR Block \(\to\) Katpadi Railway Station & 7.77 & 7.0617 & 9.12 \\
Lambert High \(\to\) Riverwatch Middle & 35.3871 & 31.6841 & 10.46 \\
\bottomrule
\end{tabular}
\caption{Runtime improvements (in seconds) across real-world routes. Each case was repeated 10 times.}
\end{table}

\section{Discussion}

\subsection{Impact of Results}

Our study underscores a subtle but meaningful performance bottleneck in Python’s use of \texttt{float('inf')} for high-frequency pathfinding tasks. Substituting with AbsInf yields consistent speedups across both synthetic and real-world graphs. Because this optimization is purely at the object level, it highlights the importance of low-level data representations in high-level interpreted languages.

\subsection{Inferences and Broader Implications}

While \texttt{float('inf')} follows IEEE-754 for broad scientific compatibility, such rigor often goes underused in algorithms like Dijkstra’s, which only require a symbolic infinity. AbsInf demonstrates that small, domain-specific optimizations can yield tangible gains even within Python’s dynamic environment. Given that time savings of 5–10\% can be significant at scale (e.g., large transportation networks or complex routing systems), these results warrant attention to specialized numeric abstractions.

\subsection{Limitations}

AbsInf is not a universal replacement for \texttt{float('inf')} in all scenarios. It is narrowly designed around comparison dominance and minimal arithmetic, omitting other floating-point behaviors (e.g., multiplication, division, or IEEE compliance). Its distribution as a compiled C extension also introduces a build step, limiting immediate portability in some environments.

\subsection{Future Work}

Potential directions include:
\begin{itemize}
\item Applying AbsInf to other graph algorithms (Bellman-Ford, A*, Floyd-Warshall).
\item Exploring similar abstractions in other languages (Java, JavaScript, C++).
\item Designing a general-purpose “symbolic constants” library for Python.
\item Investigating memory-level optimizations in adjacency matrices where “infinity” placeholders are prevalent.
\item Evaluating the impact of AbsInf in GPU-accelerated or parallelized shortest-path algorithms.
\end{itemize}

\section{Acknowledgments and AI Assistance}

This research was conducted independently and is not affiliated with, nor endorsed by, Virginia Tech or Alpharetta High School. The institutions listed are solely for author identification purposes.

The author carried out the conceptual design, implementation, and benchmarking of AbsInf. All test cases and data collection were performed manually by the author on the described hardware/cloud setups. Parts of the paper (including code and LaTeX formatting) were assisted by AI-based text-generation tools (OpenAI ChatGPT), used solely to refine clarity and structure. The core research content remain the author’s own work.

\clearpage
\appendix
\section{Appendix: Test Cases}

\subsection{Synthetic Maps}

\begin{verbatim}
# 1. Linear / Chain
Linear_Chain_1 = {"A": {"B": 2}, "B": {"C": 3}, "C": {"D": 1}, "D": {}}
Linear_Chain_2 = {"1": {"2": 5}, "2": {"3": 4}, "3": {"4": 6}, 
"4": {"5": 2}, "5": {}}

# 2. Sparse Tree
Sparse_Tree_1 = {"A": {"B": 1, "C": 2}, "B": {"D": 4},
"C": {"E": 3}, "D": {}, "E": {}}
Sparse_Tree_2 = {"1": {"2": 7}, "2": {"3": 5}, "3": {"4": 1, "5": 3},
"4": {}, "5": {}}

# 3. Dense Graph
Dense_Graph_1 = {
    "A": {"B": 2, "C": 5, "D": 1},
    "B": {"A": 2, "C": 3, "D": 2},
    "C": {"A": 5, "B": 3, "D": 1},
    "D": {"A": 1, "B": 2, "C": 1},
}
Dense_Graph_2 = {
    "1": {"2": 1, "3": 2, "4": 3},
    "2": {"1": 1, "3": 1, "4": 2},
    "3": {"1": 2, "2": 1, "4": 1},
    "4": {"1": 3, "2": 2, "3": 1},
}

# 4. Star Graph
Star_Graph_1 = {"A": {"B": 2, "C": 3, "D": 1}, "B": {}, 
"C": {}, "D": {}}
Star_Graph_2 = {"0": {"1": 5, "2": 4, "3": 6, "4": 7}, 
"1": {}, "2": {},
"3": {}, "4": {}}

# 5. Disconnected Graph
Disconnected_Graph_1 = {"A": {"B": 3}, "B": {"C": 4},
"C": {}, "X": {"Y": 1}, "Y": {}}
Disconnected_Graph_2 = {"1": {"2": 1}, "2": {}, "3": {"4": 2}, 
"4": {}, "5": {}}

# 6. Cycle Graph
Cycle_Graph_1 = {"A": {"B": 1}, "B": {"C": 2}, "C": {"A": 3}}
Cycle_Graph_2 = {"1": {"2": 1}, "2": {"3": 1}, 
"3": {"4": 1}, "4": {"1": 1}}

# 7. Equal Weights
Equal_Weights_1 = {"A": {"B": 1}, "B": {"C": 1}, "C": {"D": 1}, "D": {}}
Equal_Weights_2 = {"1": {"2": 1, "3": 1}, "2": {"4": 1}, 
"3": {"4": 1}, "4": {}}

# 8. Large Uniform Graph (3x3 grid)
Large_Uniform_Graph_1 = {
    "A": {"B": 1, "D": 1},
    "B": {"A": 1, "C": 1, "E": 1},
    "C": {"B": 1, "F": 1},
    "D": {"A": 1, "E": 1, "G": 1},
    "E": {"B": 1, "D": 1, "F": 1, "H": 1},
    "F": {"C": 1, "E": 1, "I": 1},
    "G": {"D": 1, "H": 1},
    "H": {"E": 1, "G": 1, "I": 1},
    "I": {"F": 1, "H": 1},
}
Large_Uniform_Graph_2 = {
    "1": {"2": 1, "4": 1},
    "2": {"1": 1, "3": 1, "5": 1},
    "3": {"2": 1, "6": 1},
    "4": {"1": 1, "5": 1, "7": 1},
    "5": {"2": 1, "4": 1, "6": 1, "8": 1},
    "6": {"3": 1, "5": 1, "9": 1},
    "7": {"4": 1, "8": 1},
    "8": {"5": 1, "7": 1, "9": 1},
    "9": {"6": 1, "8": 1},
}

# 9. Worst-case Tie Graph
Worst_Case_Tie_1 = {
    "A": {"B": 1, "C": 1},
    "B": {"D": 1},
    "C": {"D": 1},
    "D": {},
}
Worst_Case_Tie_2 = {
    "1": {"2": 2, "3": 2},
    "2": {"4": 2},
    "3": {"4": 2},
    "4": {},
}

# 10. Real-world-like Graph
Real_World_Like_1 = {
    "Home": {"Gas_Station": 2, "Supermarket": 5},
    "Gas_Station": {"Work": 6},
    "Supermarket": {"Work": 2},
    "Work": {},
}
Real_World_Like_2 = {
    "Apt": {"Grocery": 3, "School": 6},
    "Grocery": {"Mall": 4},
    "School": {"Mall": 2},
    "Mall": {"Office": 5},
    "Office": {},
}
\end{verbatim}

\subsection{Real-World Maps (Route Coordinates)}

\begin{verbatim}
test_routes = {
    "McComas Hall(895 Washington St SW, Blacksburg, VA 24060) To 
    Kroger South(1322 S Main St, Blacksburg, VA 24060)":
    [(37.22077736791238, -80.42247000488936),
    4000,
    (37.21689030678678, -80.40265650118901)],
    "McComas Hall(895 Washington St SW, Blacksburg, VA 24060)
    To Kroger North(903 University City Blvd, Blacksburg, VA 24060)":
    [(37.22077736791238, -80.42247000488936),
    4000,
    (37.23552264245645, -80.43524403205011)],
    "The Inn(901 Prices Fork Rd, Blacksburg, VA 24061) to Pamplin Hall
    (880 W Campus Dr, Blacksburg, VA 24061)":
    [(37.23001825689157, -80.43000178079231),
    2500,
    (37.228762507887176, -80.42473392243994)],
    "Fairfield by Marriott Bengaluru(Marathahalli - 
    Sarjapur Outer Ring Rd, Bellandur, Bengaluru, Karnataka 560103, India)
    to Decathlon Sports Sarjapura(Survey 96/1, 
    After Wipro Corporate Office - Railway Crossing, Sarjapur Main Rd, 
    Bengaluru, Karnataka 560035, India)":
    [(12.928164368027304, 77.68253489278254),
    9000,
    (12.902935714851331, 77.70700861321126)],
    "St Johns Hospital BMTC Bus Stand(WJH8+R9H, Sarjapur 
    - Marathahalli Rd, Santhosapuram, Koramangala Industrial Layout, 
    Koramangala, Bengaluru, Karnataka 560034, India) 
    to The Rameshwaram Cafe(847/1(Old No. 847/A), Binnamangala 1st Stage, 
    100 Feet Rd, Bengaluru, Karnataka 560038, India)":
    [(12.929317600267492, 77.61759068476289),
    9000,
    (12.983308622269272, 77.64085481287738)],
    "M.A. Chidambaram Stadium(1, Wallahjah Rd, Chepauk, Triplicane, 
    Chennai, Tamil Nadu 600002, India) to Chennai Lighthouse
    (Marina Beach Road,Marina Beach, Mylapore, Chennai,
    Tamil Nadu 600004, India)":
    [(13.063095632837598, 80.27942274907355),
    4000,
    (13.040466160210698, 80.27929182736985)],
    "Dr. MGR Block(X594+J88, Vellore, Tamil Nadu 632014, India) 
    to Katpadi Railway Station(Katpadi Jct, KRS Nagar, Katpadi, 
    Vellore, Tamil Nadu 632007, India)":
    [(12.969217680093521, 79.15580660978962),
    4000,
    (12.972139845854452, 79.13782641820697)],
    "Lambert High School(805 Nichols Rd, Suwanee, GA 30024) 
    to Riverwatch Middle School(610 James Burgess Rd #1135, 
    Suwanee, GA 30024)":
    [(34.10641389262153, -84.13890857432139),
    7000,
    (34.12447446253152, -84.10939258191553)]
}
\end{verbatim}

\end{document}